\documentclass[11pt]{article}

\newcommand{\beq}{\begin{equation}}
\newcommand{\eeq}{\end{equation}}
\newcommand{\p}{\partial}

\begin{document}

\vspace*{1.in}

\begin{center}

{\Large {\bf Einstein's Violation of General Covariance}} \\

\vspace{.5in}

{\large {\bf Kenneth Dalton}} \\

\vspace{.2in}

email: kxdalton@yahoo.com

\vspace{1.5in}

{\bf Abstract}

\end{center}

\vspace{.25in}

Einstein rejected the differential law of energy-momentum
conservation $ T^{\mu\nu}_{;\nu} = 0 $.  In doing so, he
violated the principle of general covariance.  Here, we
prove the conservation law $\, T^{\mu\nu}_{;\nu} = 0 \,$
and discuss its significance for general
relativity.

\clearpage

In his founding paper on general relativity, Einstein stated
the principle of general covariance:

\begin{quotation}

``The general laws of nature are to be expressed by 
equations which hold good for all systems of coordinates, 
that is, are covariant with respect to any substitutions
whatever (generally covariant).'' [1]

\end{quotation}
In that same paper, he rejected the differential law of 
energy-momentum conservation

\beq
      T^{\mu\nu}_{;\nu} =
             \frac{1}{\sqrt{-g}}
             \frac{\p \sqrt{-g}\,\, T^{\mu\nu}}{\p x^\nu}
               + \Gamma^\mu_{\nu\lambda} T^{\nu\lambda}
                = 0
\eeq
He writes as follows: the term $ \Gamma^\mu_{\nu\lambda}
T^{\nu\lambda} $ ``shows that laws of conservation of
momentum and energy do not apply in the strict sense
for matter alone.''[2]  This statement constitutes a direct
violation of general covariance.  All textbook authors to
date have followed Einstein's erroneous lead in this matter.
Here are two exceptionally forthright quotations:

\begin{itemize}

\item[(a)] the equations $ T^{\mu\nu}_{;\nu} = 0 $ ``are not
what can properly be called conservation laws'' [3];

\item[(b)] the equation $ T^{\mu\nu}_{;\nu} = 0 $ ``does not
generally express any conservation law whatever.'' [4]

\end{itemize}

It is not difficult to prove the conservation law (1).  In flat
rectangular coordinates $ x^\mu = (x^0, x, y, z) $
conservation is expressed by the Lorentz covariant equation

\beq
         \frac{\p \,T^{\mu\nu} }{\p x^\nu} = 0
\eeq
Suppose, instead, that we choose ordinary flat spherical
coordinates $ {x^\mu} ' = (x^0, r, \theta , \phi ) $.  What
will be the correct equation in the new coordinate system?
To answer this question, we begin with equation (2) and 
substitute the transformed quantities 

\beq
    T^{\mu\nu} = \frac{\p x^\mu}{\p {x^\alpha}'}
                       \frac{\p x^\nu}{\p {x^\beta}'}\,
                        {T^{\alpha\beta}}'
\eeq

\beq
     \frac{\p }{\p x^\nu} = \frac{\p {x^\gamma}'}{\p x^\nu}
                                  \frac{\p }{\p {x^\gamma}'}
\eeq
We then make use of 

\beq
      {\Gamma^\mu_{\nu\lambda}}'
        = \frac{\p {x^\mu}'}{\p x^\alpha}
          \frac{\p x^\beta}{\p {x^\nu}'}
          \frac{\p x^\gamma}{\p {x^\lambda}'}\,
          \Gamma^\alpha_{\beta\gamma}
          + \frac{\p {x^\mu}'}{\p x^\alpha}
            \frac{\p^2 x^\alpha}{\p {x^\nu}' \p {x^\lambda}'}
\eeq

\beq
     \frac{1}{\sqrt{-g}\,'} \frac{\p \sqrt{-g}\,'}{\p {x^\nu}'}
              = {\Gamma^\alpha_{\alpha\nu}}'
\eeq
and arrive at the equation

\beq
     \frac{1}{\sqrt{-g}\,'}
     \frac{\p \sqrt{-g}\,'\, {T^{\mu\nu}}'}{\p {x^\nu}'}
          + {\Gamma^\mu_{\nu\lambda}}'\, {T^{\nu\lambda}}'
               = 0
\eeq
This proves the differential law of energy-momentum
conservation in the spherical coordinate system.
Because this law is generally covariant, it must hold
true for all systems of coordinates, flat or curved 
(principle of general covariance).

The equation $ \, T^{\mu\nu}_{;\nu} = 0 \, $ 
is not open to interpretation, any more than Maxwell's equations 
are open to interpretation.  They are generally covariant laws
of nature, all of which belong to four-dimensional tensor analysis.
These laws are {\it beyond personal choice}; it is this 
fact which demonstrates the power of general covariance.
To reject such an equation is simply to make a mistake.
Einstein made such a mistake in rejecting the law of
energy-momentum conservation $\,T^{\mu\nu}_{;\nu} = 0.$

This law has profound physical consequences for general
relativity.  The gravitational field equations are

\beq
      R^{\mu\nu} - \frac{1}{2} g^{\mu\nu} R
                    = - \kappa \, T^{\mu\nu}
\eeq
$ T^{\mu\nu} $ is the stress-energy-momentum tensor
of matter and electromagnetism.  The covariant
divergence of the left-hand side is identically zero,
therefore

\beq
      T^{\mu\nu}_{;\nu} = 0
\eeq
This equation means that the energy-momentum of matter
and electromagnetism is conserved, at all space-time
points.  In other words, there is no exchange of 
energy-momentum with the gravitational field.  Conclusion:
Einstein's gravitational field has no energy, momentum, 
or stress [5, 6].

\clearpage

\section*{\large {\bf References}}

\begin{enumerate}

\item A. Einstein, ``The Foundation of the General Theory
         of Relativity'' in {\it The Principle of Relativity}
         (Dover, New York, 1952) section 3.
\item A. Einstein, Ref. 1, section 18.
\item P. Bergmann, {\it Introduction to the Theory of
         Relativity} (Dover, New York, 1976) page 194.
\item L. Landau and E. Lifshitz, {\it The Classical Theory
         of Fields} (Pergamon, Elmsford, 1975) 4th ed.,
         section 96.
\item K. Dalton, ``Energy and Momentum in General Relativity'',
         {\it Gen. Rel. Grav.} {\bf 21}, 533-544 (1989).
\item K. Dalton, ``Manifestly Covariant Relativity'', {\it Hadronic J.} {\bf 17}, 139-142 (1994); 
         also, www.arxiv.org/physics/0608030.

\end{enumerate}

\end{document}